\documentclass[showpacs, prb,twocolumn,preprintnumbers , superscriptaddress, aps]{revtex4-2}

\usepackage{color}
\usepackage{amsmath,amssymb}
\usepackage{pifont}
\usepackage{amssymb}  
\usepackage{bbold}
\usepackage{float}
\usepackage{subfloat}

\usepackage[labelfont=bf]{caption} 
\usepackage{tikz}
\usepackage{makecell}
\usepackage{subfig}
\usepackage{pifont}   
\usepackage{graphicx} 
\usepackage{dcolumn}  
\usepackage{bm}       
\usepackage{multirow} 
\usepackage{placeins}
\usepackage[colorlinks]{hyperref}
\usepackage{mathtools}

\newcommand{\norm}[1]{\left\|{#1}\right\|}

\usepackage{breqn}
\usepackage{mathrsfs}

\usepackage{multirow}

\begin{document}
\title{Linear differential equation approach to the Loschmidt amplitude}
\author{Michael Vogl}
		\affiliation{Physics Department$,$
		King Fahd University
		of Petroleum $\&$ Minerals$,$
		Dhahran 31261$,$ Saudi Arabia}
    \affiliation{Interdisciplinary Research Center (IRC) for Intelligent Secure Systems$,$ KFUPM$,$ Dhahran$,$ Saudi Arabia}

  \begin{abstract}
The Loschmidt amplitude is a popular quantity that allows making predictions about the stability of quantum states under time evolution. In our work, we present an approach that allows us to find a differential equation that can be used to compute the Loschmidt amplitude. This approach, while in essence perturbative, has the advantage that it converges at finite order. We demonstrate that the approach for generically chosen matrix Hamiltonians often offers advantages over Taylor and cumulant expansions even when we truncate at finite order. We then apply the approach to two ordinary band Hamiltonians (multi-Weyl semimetals and AB bilayer graphene) to obtain the Loschmidt amplitude after a quench for an arbitrary starting state and find that the results readily generalize to find transmission amplitudes and specific contributions to the partition function, too. We then test our methods on many body spin and fermionic Hamiltonians and find that while the approach still offers advantages, more care has to be taken than in a generic case. We also provide an estimate for a breakdown time of the approximation.
	\end{abstract}
\maketitle

 \section{Introduction}
With current rapid developments in quantum computing, it has increasingly become more crucial to make accurate predictions about the stability of quantum states. For instance, it is beneficial to predict the stability of quantum bits \cite{Burnett_2019,PhysRevLett.128.127702,PhysRevB.109.144502} because it ensures the reliability of quantum computations. The stability of the underlying ground state that plays host to quantum bits is also essential because it ensures that quantum bits can properly be generated \cite{roberts2022fidelity}. One measure of state stability is the so-called Loschmidt amplitude\cite{GORIN200633,Heyl_2018}, which measures the overlap between states at different times in their evolution. However, it cannot be overstated that the utility of this quantity reaches far beyond this application. Indeed, it has been found that it can be used to define the term of a dynamical phase transition \cite{Heyl_2018,vzunkovivc2016dynamical} using its discontinuities.\\
While computing the overlap between two states seems conceptually trivial, it is essential to notice that it is plagued by the difficulty that time evolution involves a matrix exponential of the Hamiltonian - such a quantity is generally difficult to compute when matrices grow in size\cite{doi:10.1137/S00361445024180}. On the face of it, it may seem odd that computing a scalar quantity's time evolution like the Loschmidt amplitude involves working with a high dimensional Hilbert space. One may wonder why, so far, it is not possible just to identify initial conditions and then evolve the Loschmidt amlitude according to those initial conditions without ever making further reference to a large Hilbert space. This work aims to achieve the goal of finding both appropriate initial conditions and the corresponding evolution equations. 

To achieve this goal, we structured our work as follows. In Sec. \ref{sec:2}, we summarize the more typical approximate approaches to compute the Loschmidt amplitude. Sec. \ref{sec:3} demonstrates how differential equations for the Loschmidt amplitude can be found. In Sec. \ref{sec:4}, we then analyze the properties and shortcomings of the approach. The section begins by testing the approach on thousands of randomly chosen matrix Hamiltonians, which permits us to understand the generic properties of the approach. We test the approach on two band Hamiltonians where we obtain readily generalizable expressions for the Loschmidt amplitude of a multi-Weyl semi-metal and AB bilayer graphene. Lastly, we test the approach on two common many-body toy models (a spin model and a fermionic model) and find that the approach offers advantages over standard approximation methods. However, proper care has to be taken. Finally, in Sec. \ref{sec:5}, we draw conclusions and state what follow-up work might be interesting.
 
 \section{Loschmidt amplitude and common computational schemes}
 \label{sec:2}
We begin our discussion by noting that any quantum state $\left| \psi(t)\right\rangle$ evolved by some Hamiltonian $\hat H$ will start to deviate from its original state $\left| \psi_0\right\rangle$. This deviation can be quantified by the so-called Loschmidt amplitude \cite{Heyl_2018}
\begin{equation}
G(t)=\left\langle\psi_0 \mid \psi(t)\right\rangle=\left\langle\psi_0\left|e^{-i H t}\right| \psi_0\right\rangle,
\label{eq:Loschmidt_amplitude}
\end{equation}
which, as noted by Heyl \cite{Heyl_2018,Heyl_2019survey} carries strong formal similarities to the partition function in statistical mechanics (one does a Wick-rotation $t\to-i\beta$ and replaces the expectation value by a trace $\langle\psi_0|\dots|\psi_0\rangle \to \operatorname{tr}(\dots)$). Because of this similarity, it is commonly used to diagnose dynamical phase transitions \cite{Heyl_2018,Heyl_2019survey}. In addition to its utility for dynamical phase transitions, it also has proven its usefulness in our understanding of stability in quantum computing, where it is closely related to the fidelity of a quantum state\cite{Tonielli_2019,jacquod2009decoherence}.\\

Now typically, as we can see from our expression \eqref{eq:Loschmidt_amplitude} computations of the Loschmidt amplitude often require being able to i) determine operator exponentials $\exp(-i H t)$ and ii) compute their expectation values. Such operator exponentials are often difficult to obtain - especially in the many-body context, such as for fermionic interacting model and spin Hamiltonians, where the Hilbert space can be excessively large. Moreover, even the expectation value for a generic state $\psi_0$ can also be cumbersome to compute. A typical simplistic approach to side-step this problem is to work with a series expansion that is valid for small times
\begin{equation}
    G(t)\approx \sum_{n=0}^{n^*}\frac{(-it)^n}{n!}\left\langle\psi_0\left|H^n\right| \psi_0\right\rangle,
    \label{eq:Taylor}
\end{equation}
where $n^*$ is some cut-off order. This approach, of course, has the drawback that $t^{n^*}$ will grow without bounds, limiting the range of validity of the approach. Moreover, computing $\left\langle\psi_0\left|H^n\right| \psi_0\right\rangle$ becomes exceedingly difficult as $n$ grows. 

The first issue of boundless growth can be addressed by the more clever idea of a cumulant expansion, where one takes
\begin{equation}
    G(t)\approx \exp\left(\sum_{n=1}^{n^*}\frac{(-it)^n}{n!}C_n(H)\right),
    \label{eq:Cumulant}
\end{equation}
and $C_n(H)$ refers to the $n$-th cummulant or connected average of $H$ (i.e. $C_1(H)=\left\langle\psi_0\left|H\right| \psi_0\right\rangle$ and $C_2(H)=\left\langle\psi_0\left|H^2\right| \psi_0\right\rangle-\left\langle\psi_0\left|H\right| \psi_0\right\rangle^2$ etc.). Here, the boundless growth is prevented by the exponential form. 

Of course, as we will see later in both cases, it is very difficult to obtain results that are valid for a very long time, and we want to improve upon this issue in certain cases.

We note that, of course, in cases where the Hamiltonian $H=H_0+V$ with a perturbation $V$ that can be considered as small and $H_0$ that has an easily computable exponential $e^{-iH_0t}$ one can improve on these approaches - for instance by use of an interaction picture \cite{Tonielli_2019} and obtain results valid for long times. However, we aim to address the generic case where this is not always possible. An improvement on typical techniques in the interaction picture can be the topic of future work.

 \section{Proposed scheme}
 \label{sec:3}
Our goal is to derive approximate linear differential equations for the Loschmidt amplitude $G(t)$ that lead to approximate results that improve on the common methods discussed earlier without adding more complications. Finding differential equations for $G(t)$ by itself is challenging. For instance, we may start with the Schr\"odinger equation $i\partial_t |\psi(t)\rangle=H|\psi(t)\rangle$ and multiply on the right with $\left\langle\psi_0\right|$ to obtain
\begin{equation}
    i\partial_t G(t)=\left\langle\psi_0\right|H|\psi(t)\rangle,
\end{equation}
which however involves not just $G(t)$ but also the matrix element $\left\langle\psi_0\right|H|\psi(t)\rangle$. This matrix element is difficult to relate to derivatives $\partial_t^nG(t)$ in a useful fashion. Indeed, even if we use the identity
\begin{equation}
    \left\langle\psi_0\right|H^n|\psi(t)\rangle=i^n\partial_t^nG(t)
\end{equation}
for the only easily applicable case $n=1$ we end with a trivial tautology $i\partial_tG(t)=i\partial_tG(t)$, which is not useful. However, it can be noted that a linear function $x$ on a finite interval $[a,b]$ can be approximated as $x\approx c_0+c_2x^2+c_3x^3+c_4x^4+\dots$ to arbitrary precision if we choose appropriate constants $c_i$. The same, of course, would be true for a bounded Hamiltonian that could be approximated as
\begin{equation*}
    H\approx H_{\mathrm{approx}} =c_0\mathbb{1}+\sum_{n=2}^{n^*}c_nH^n.
\end{equation*}
This expression leads to improved approximations as the cutoff order $n^*$ increases. Evidently, this series is controlled if $H$ is small (or equivalently, the time evolution is for short times) - similar to the two other approximate series we introduced and will use as a benchmark. The difference here is that we will introduce differential equations that decide on the form of the Loschmidt amplitude $G(t)$ instead of making an apriori assumption (bare power series or power series inside of an exponential). In such a case, one would find a differential equation for the Loschmidt amplitude 
\begin{equation}
    i\partial_tG(t)=c_0 G(t)+\sum_{n=2}^{n^*}i^n c_n\partial_t^n G(t),
    \label{eq:Diffeq}
\end{equation}
which for a cut-off $n^*$ requires $n^*-1$ initial conditions
\begin{equation}
    \Bigg\{G(0),\left.\partial_tG(t)\right|_{t=0},\dots,\left.\partial_t^{n^*-1}G(t)\right|_{t=0}\Bigg\}
\end{equation}
with
\begin{equation}
    \left.\partial_t^{n}G(t)\right|_{t=0}=\left\langle\psi_0\left|H^n\right| \psi_0\right\rangle.
\end{equation}
It is noteworthy that this scheme does not pose major additional complications over the expansions previously discussed. Once we have obtained constants $c_n$, the main complication comes from setting up the initial conditions, which correspond to the same expectation values that are needed in the other two schemes.

Therefore, methods to obtain the constants $c_n$ remain to be determined.

We expect the most reliable approach for obtaining the constants $c_n$ is via a variational principle with a cost function. The most obvious such cost function, which is also easiest to work with because it is quadratic in $c_n$, is
 \begin{equation}
C=\norm{H_{\mathrm{approx.}}-H}_F^2=\mathrm{tr}[(H_{\mathrm{approx.}}-H)^2],
\end{equation}
which measures how closely the Hamiltonian is approximated.

That is, one has to solve the following set of equations
\begin{equation}
    c_0\mathrm{tr}(H^m)+\sum_{n=2}^{n^*}c_n\mathrm{tr}(H^{n+m})-\mathrm{tr}(H^{m+1})=0
    \label{eq:fitparam}
\end{equation}
for $c_i$, where $m\in\{0,2,3\dots,n^*\}$.\\
It is important to note that the involved traces often do not lead to any considerable complications because computing traces of operators is often considerably easier than computing expectation values. For instance in the case of spin operators $S^x_i$, $S^y_i$ and $S^z_i$ one has 
\begin{equation}
    \mathrm{tr}(\prod_i S^{k_i}_i)=0
\end{equation}
such that only spin matrices that multiply to be proportional to an identity contribute to a trace. This observation makes it possible to compute traces even in cases far beyond the reach of exact diagonalization algorithms. 
An interesting but obvious thing to note about this approach is that for a Hamiltonian of dimension $n$, a cutoff of $n^*=n-1$ already gives an exact result in a generic case. This observation is seen most readily because a Hamiltonian in its diagonal basis only has $n$ entries that the variational principle must match. In a typical generic case, $n^*=n-1$ means that we have $n$ constants to match and, therefore, can match every single eigenvalue in the trace.\\
It is now clear that the approach, unlike the case for a cumulant expansion or a Taylor series, becomes exact at finite order. We will see in numerical experiments that it is an improvement on those approaches also at small orders.

We should note that an exact result at finite order means that the series is convergent for a finite size Hilbert space, which makes it possible to study convergence acceleration techniques. We leave this topic for a later study.
 
 \section{Results and properties}
 \label{sec:4}
 \subsection{Generic properties}
As a first example to test our theory, we begin by considering randomly chosen hermitian matrices and randomly chosen evolved vectors. Here, one should keep in mind that we want to compare results for similar levels of difficulty in computation. That means we want to compare results that involve the same order of $n$ of expectation value $\langle H^n\rangle$. We will compare the Loschmidt echo for a particular case (the case where it is the same as the so-called survival probability) for various approximations. For simplicity in the remainder of the text, we will refer to it simply as Loschmidt echo but we caution the reader to have in mind that this is particular case of the Loschmidt echo - not the most general case. The expression for the particular case is given below
\begin{equation}
    L(t)=\left|G(t)\right|^2
\end{equation}
directly for the different approximations, and we will also look at plots of the relative approximation error
\begin{equation}
    \Delta L_{\mathrm{rel}}(t)=\left|\frac{L(t)-L_{\mathrm{approx}}(t)}{L(t)}\right|,
\end{equation}
which measures the relative error between exact Loschmidt echo $L(t)$ and approximate Loschmidt echo $L_{\mathrm{approx}}(t)$.

Below in Fig. \ref{fig:f1} we compare various results for randomly chosen $3\times 3$ Hamiltonians $H$.
\begin{figure*}[htbp!]
    \centering
    \includegraphics[width=1\linewidth]{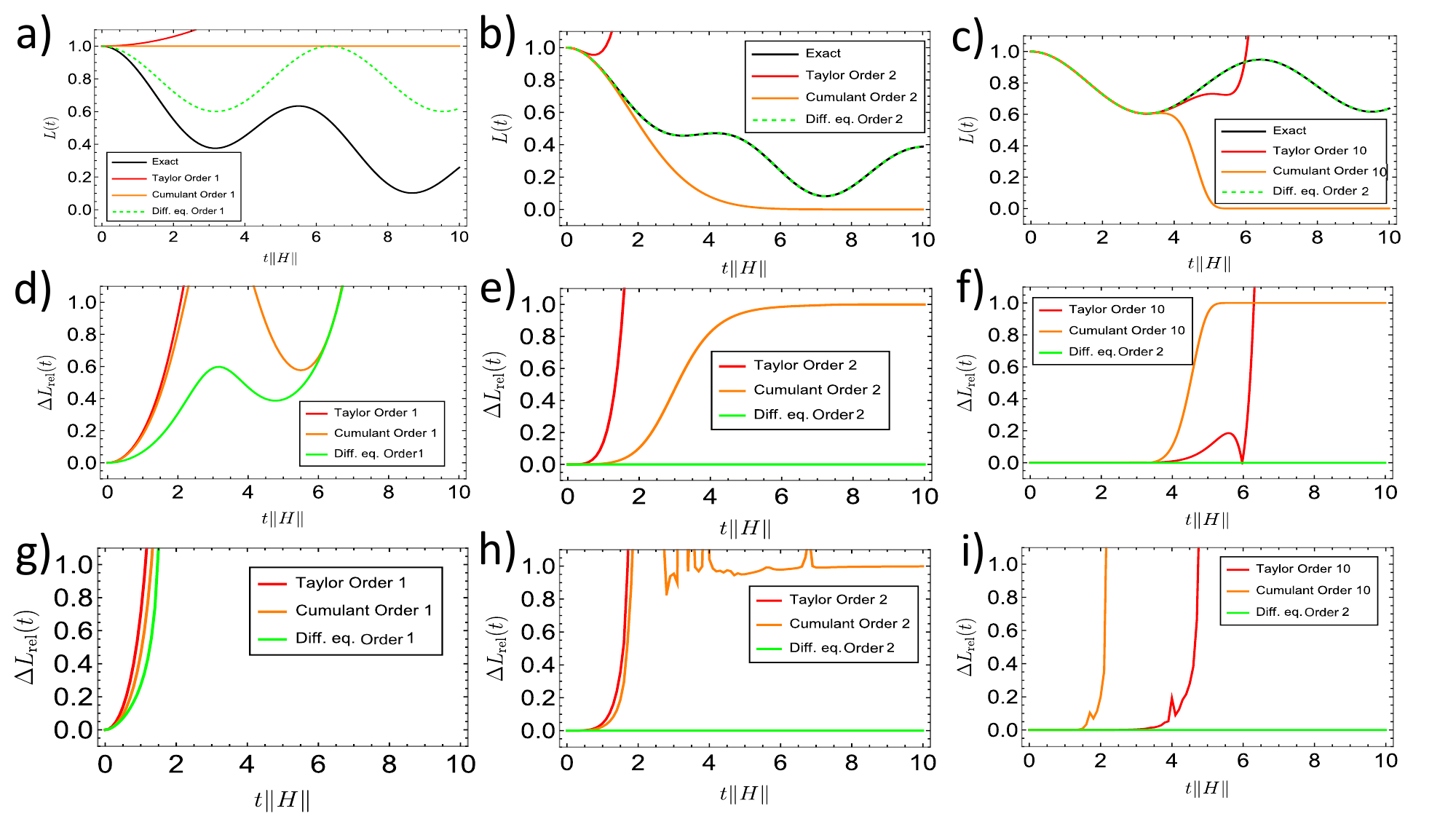}
    \caption{In this figure, we present plots of the Loschmidt echo for different realizations of $3\times 3$ Hamiltonians. In the first row (subfigures (a-c)), we see representative example plots of the Loschmidt echo $L(t)$. In the second row (subfigures (d-f)), we see representative example plots of the relative error for Loschmidt echo $\Delta L_{\mathrm{rel}}(t)$. We plotted the relative error that was averaged over 1000 realizations in the last row. The first column (a,d,g) corresponds to approximations of order 1, the second column (b,e,h) to approximations of order 2, and the last column (c,f,i) to approximations of order 10. All subfigure plots are a function of dimensionless time $t\lVert H\rVert$ to ensure an easy way to compare different Hamiltonians $H$, where the spectral radius norm is used. In all plots, the exact result \eqref{eq:Loschmidt_amplitude} is in black, the Taylor series result (Eq. \eqref{eq:Taylor}) in red, the result from a cumulant series (Eq. \eqref{eq:Cumulant}) in orange and the result from the differential equation approach (Eq. \eqref{eq:Diffeq}) in green.}
    \label{fig:f1}
\end{figure*}
As mentioned in the previous section, for $3\times 3$ Hamiltonians, one expects order $2$ results from the differential equation approach to give exact results, and this is indeed the case as one can instantly see from our plots (Fig.\ref{fig:f1}(b,c,e,f,h,i)). Indeed, as expected, this is generically the case even for 1000s of randomly chosen Hamiltonians(Fig.\ref{fig:f1}(h,i)). Another thing to notice from our computations is that unlike the Taylor series and the cumulant expansion, our differential equation approach can sometimes capture qualitative features of the Loschmidt echo even at order 1. 

\begin{figure*}[htbp!]
    \centering
    \includegraphics[width=1\linewidth]{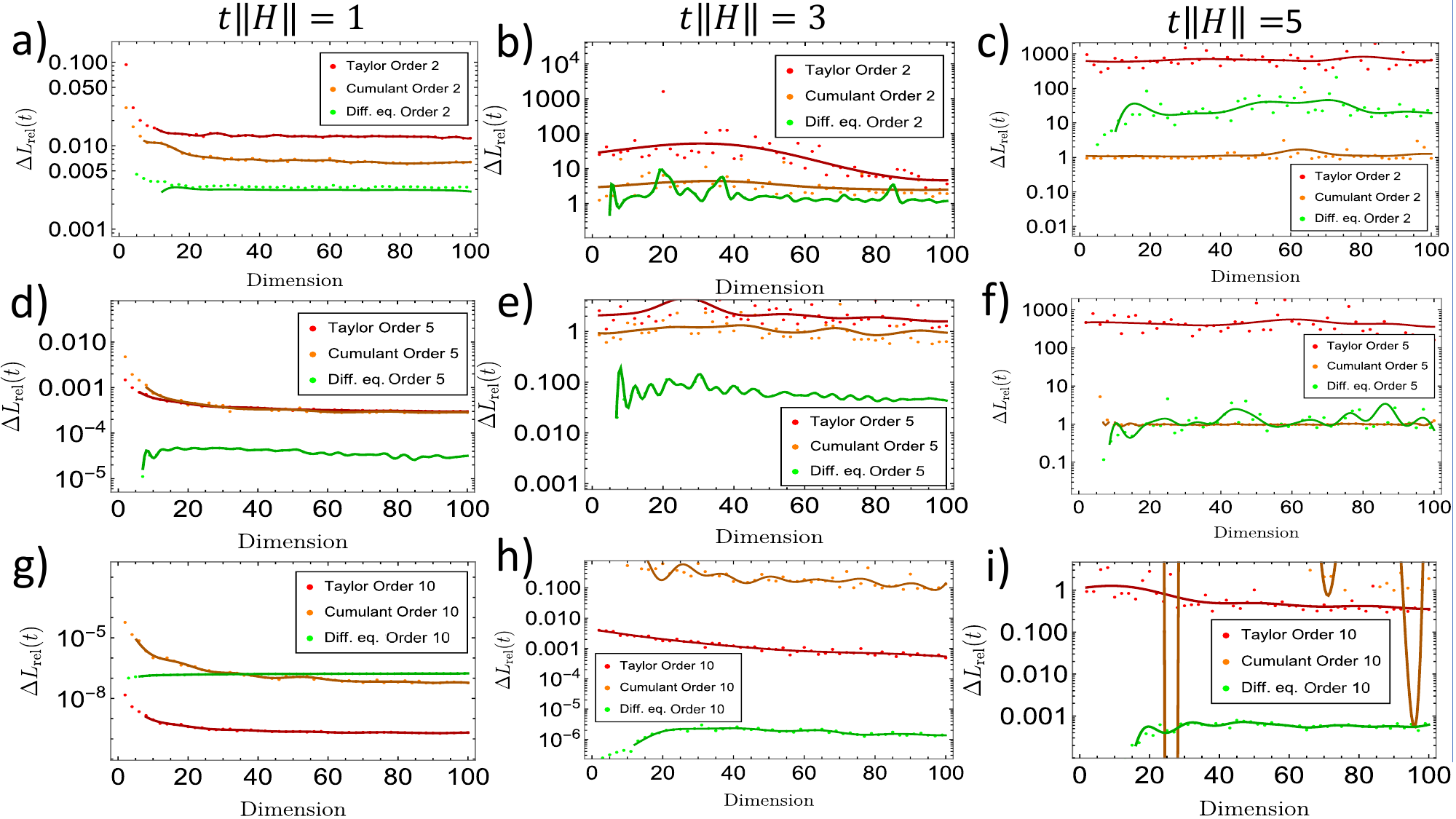}
    \caption{Here, we present plots of the relative approximation error for the Loschmidt echo $\Delta L_{\mathrm{rel}}$ as a function of Hilbert space dimension. The different rows are associated with different approximation orders, with the first row (subfigures (a-c)) corresponding to order 2, the second row (subfigures (d-f)) to order 5, and the last row (subfigures (g-i)) to order 10. The different columns correspond to different slices of dimensionless time $t\lVert H\rVert$ with left column $t\lVert H\rVert=1$, middle column $t\lVert H\rVert=3$ and the right column $t\lVert H\rVert=5$. Furthermore, every data point is averaged over 200 realizations to ensure that results are as generic as possible. The Taylor series result (Eq. \eqref{eq:Taylor}) is plotted in red, the result from a cumulant series (Eq. \eqref{eq:Cumulant}) in orange, and the result from the differential equation approach (Eq. \eqref{eq:Diffeq}) in green. To guide the reader's vision, we have included slightly darker color solid lines that were obtained by a Gaussian process.}
    \label{fig:f2}
\end{figure*}

A relatively natural next question is whether or not the approximation stays reliable if the dimension of the Hilbert space is increased. To understand this, we have generated various plots that visualize approximation errors as a function of Hilbert space dimension, which can be seen in Fig.\ref{fig:f2} 
From the plots in Fig. \ref{fig:f2}, it becomes clear that the validity of the various approximations does not crucially depend on the size of the Hilbert space. If anything, one might see a slight improvement in approximation error in larger matrix sizes. This statement can be confirmed at different approximation orders (the different rows of the figure) and at different points in time (the different columns). We also notice that in virtually all cases, the approximation in the differential equation approach we advocate here is several orders of magnitude more reliable than the usual cumulant expansion or Taylor series approaches.

Of course, it is interesting to see how this looks in detail for relatively large Hamiltonians of size $200\times 200$, which we show below in Fig. \ref{fig:f3}.
\begin{figure*}[htbp!]
    \centering
    \includegraphics[width=1\linewidth]{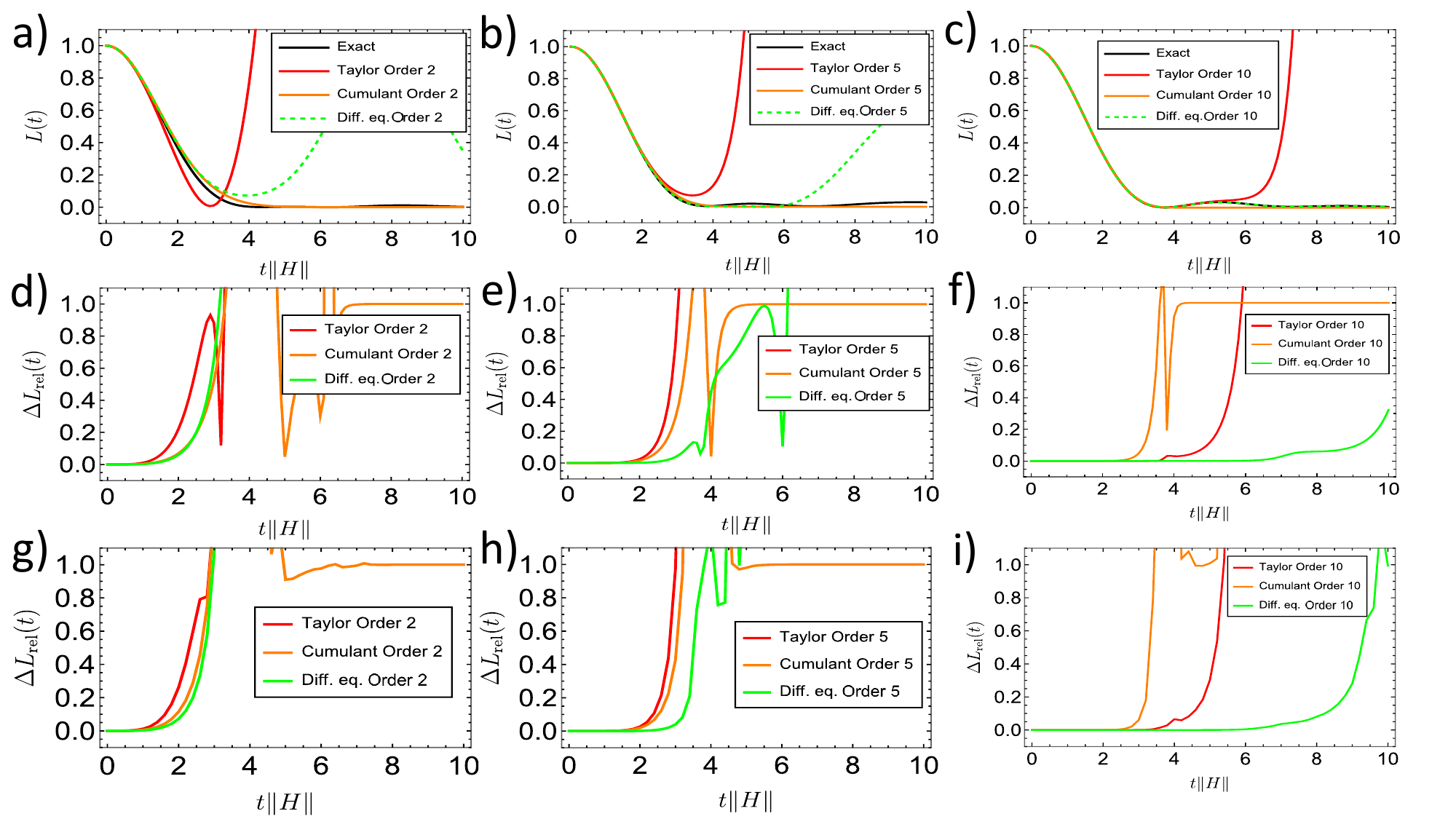}
    \caption{In this figure, we present plots of the Loschmidt echo for different realizations of $200\times 200$ HamiltoniansIn the first row (subfigures (a-c)), we see representative example plots of the Loschmidt echo $L(t)$. In the second row (subfigures (d-f)), we see representative example plots of the relative error for Loschmidt echo $\Delta L_{\mathrm{rel}}(t)$. We plotted the relative error, which was averaged over 1000 realizations in the last row. The first column (a,d,g) corresponds to approximations of order 1, the second column (b,e,h) to approximations of order 2, and the last column (c,f,i) to approximations of order 10. All subfigure plots are a function of dimensionless time $t\lVert H\rVert$ to ensure an easy way to compare different Hamiltonians $H$, where the spectral radius norm is used. In all plots, the exact result \eqref{eq:Loschmidt_amplitude} is plotted in black, the Taylor series result (Eq. \eqref{eq:Taylor}) in red, the result from a cumulant series (Eq. \eqref{eq:Cumulant}) in orange and the result from the differential equation approach (Eq. \eqref{eq:Diffeq}) in green.}
    \label{fig:f3}
\end{figure*}
From Fig. \ref{fig:f3} in the upper row (subfigures (a-c)), it becomes clear that the differential equation approach is much better than both Cumulant and Taylor series expansions at capturing partial revivals. Furthermore, in the bottom two rows, we see that the differential equation approach advocated here often is valid for up to twice as long dimensionless times $t\lVert H\rVert$. 

\begin{figure}[htbp!]
    \centering
    \includegraphics[width=1\linewidth]{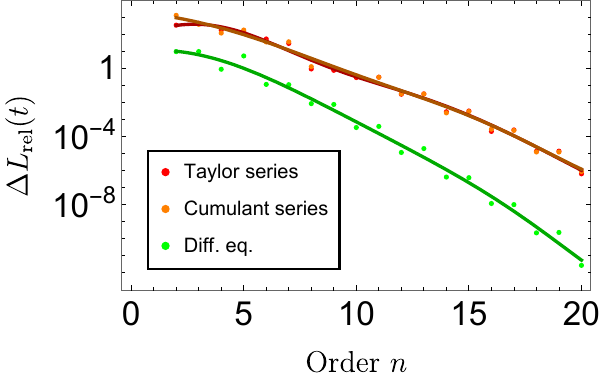}
    \caption{Here we present plots of the relative approximation error for the Loschmidt echo $\Delta L_{\mathrm{rel}}$ as a function of approximation order $n$. Because results, as previously seen, should be relatively independent of Hilbert space dimensionality, we consider here the case of $200\times 200$ Hamiltonians; the plot is taken at a dimensionless time $t\lVert H\rVert=5$. The Taylor series result (Eq. \eqref{eq:Taylor}) is plotted in red, the result from a cumulant series (Eq. \eqref{eq:Cumulant}) in orange, and the result from the differential equation approach (Eq. \eqref{eq:Diffeq}) in green. To guide the reader's vision, we have included solid lines in slightly darker colors that were obtained by a Gaussian process. We note that every data point has been averaged over 100 realizations of the Hamiltonian.}
    \label{fig:f4}
\end{figure}

Last, we are interested to learn more generally about convergence properties of the different series, which can be understood from Fig. \ref{fig:f4}
From Fig. \ref{fig:f4}, we see that all different approximations roughly converge exponentially fast (almost straight line in a logarithmic plot). However, the differential equation approach is several orders of magnitude more reliable and seems to converge with a larger exponent. The results for the Cumulant expansion and Taylor series, on the other hand, do not qualitatively differ from one another.
 \subsection{Behaviour for examples from condensed matter theory}
 Understanding which approximations are valid in a generic setting is essential. However, as physicists, we are most interested in understanding whether an approximation case can fruitfully be applied to the types of Hamiltonians that appear in physics. We have chosen to restrict our discussion to some types of Hamiltonians relevant to condensed matter theory. 
 \subsubsection{Band Hamiltonians and exactly solvable cases}
 Relatively typical Hamiltonians that in the condensed matter context come to mind first are band Hamiltonians such as they appear in the context of graphene systems \cite{Castro_Neto_2009,McCann_2013,Min_2008} or for semi-metals \cite{PhysRevX.7.021019,annurev:/content/journals/10.1146/annurev-conmatphys-031016-025458,Bouhlal_2022,PhysRevLett.108.140405,PhysRevResearch.6.013048,Fang_2012,BOUHLAL2021168563}. Since those types of Hamiltonians typically break into momentum sectors of finite - often small - dimension, we know that the differential equation approach to the Loschmidt amplitude given in this work can produce exact results even at low orders. For instance, this would be the case for the Hamiltonian of a multi-Weyl semi-metal\cite{Fang_2012}
 \begin{equation}
     H=(ak_+^n+bk_-^n)\sigma_++h.c.+\lambda k_z\sigma_z+E_0\mathbb{1}.
 \end{equation}
Here, each k-value has a block that is only $2\times2$ ( with a vector $(1,0)$ corresponding to a valence band and $(0,1)$ to a conduction band state). The matrices $\sigma_\pm$ and $\sigma_z$ are typical Pauli matrices and $a,b,\lambda$ numerical constants. The parameter $n\in \mathbb{N}$ characterizes the type of multi-Weyl semi-metal(the order of band touchings). For the full class of such Hamiltonians, the first-order expansion would already give exact results for the Loschmidt amplitude. Since it is instructive, we will go through the example. First, one notices from equation \eqref{eq:fitparam} that fit parameters when we express the Hamiltonian in a form $c_1\mathbb{1}+c_2H^2$ are given as
\begin{equation}
    c_1=-\frac{\left| a k_+^n+b k_-^n\right| ^2-E_0^2+\lambda ^2 k_z^2}{2 E_0};\quad
    c_2=\frac{1}{2E_0}.
\end{equation}
At this point, one may be worried about the appearance of $E_0$ in the denominator in case $E_0\to0$ might be badly behaved. This worry, however, is not an actual problem, as we will see.\\
Of course, one next ends up with the differential equation
\begin{equation}
\begin{aligned}
    &i\partial_t G(t)=c_1G(t)-c_2\partial_t^2G(t)\\ &G(0)=\langle H^0\rangle=1;\quad i\partial_tG(0)=\langle H^1\rangle,
    \end{aligned}
    \end{equation}
    which has a simple solution
\begin{equation}
    \begin{aligned}
        G(t)&=e^{-\frac{i t}{2 c_2}} \left[\langle H^0\rangle \left(\cos ( \omega t)+\frac{i \sin ( \omega t)}{2 c_2 \omega }\right)-i\langle H^1\rangle\frac{ \sin (\omega t)}{\omega }\right]\\
        \omega&=\frac{\sqrt{1-4 c_1 c_2}}{2 c_2}.
    \end{aligned}
\end{equation}
Now, as we have claimed previously, the case of $E_0\to0$ is not a problem, and we obtain
\begin{equation}
    G(t)=G_0\cos (\omega_0  t)-i\frac{ \langle H\rangle \sin (\omega_0  t)}{\omega_0 },
\end{equation}
where we have used that in this limit $c_2^{-1}\to 0$ and $\omega\to\omega_0=\sqrt{\lambda^2k_z^2+\left| a k_+^n+b k_-^n\right| ^2}$.

It is important to stress that the approach advocated in this paper makes it easier to separate the equation into short pieces and makes it very clear where $\langle H\rangle=\left\langle\psi\right|H\left|\psi\right\rangle$ enters the expression - as an initial condition. Blindly taking a matrix exponential followed by an expectation value leaves this less apparent. It is interesting to note that the equation we derived can also be used to compute transition amplitudes between two states $\phi$ and $\psi$ if we replace $\langle H^n\rangle\to\left\langle\phi\right|H^n\left|\psi\right\rangle$. Moreover, after a Wick-Rotation and replacing $\langle H^n\rangle\to \operatorname{tr}(H^n)$, it can also be employed to compute the single-particle fixed momentum contribution to the partition function (for a full many-body Hamiltonian similar replacements can be made to find the full partition function).

Another simple example is AB stacked n-layer graphene, which we describe by a $2n\times 2n$ Hamiltonian such that the $2n-1$th order of the differential equation approach gives exact results. Here, we consider the example of the Hamiltonian for AB bilayer graphene, which is given below\cite{McCann_2013}
 \begin{equation}
     H=v(\mathbb{1}\otimes \sigma_x k_x+\mathbb{1}\otimes \sigma_y k_y)+\frac{t}{2}(\sigma_x\otimes \sigma_x -\sigma_y\otimes \sigma_y )+E_0\mathbb{1},
 \end{equation}
 where $\sigma_i$ are Pauli matrices, the first term in each Kronecker product corresponds to a layer degree of freedom and the second term to a sublattice degree of freedom. The constant $v$ is the Fermi velocity, and $t$ is the hopping amplitude for interlayer hopping.
 Here, it is also straightforward to find a differential equation. In the limit $E_0\to 0$, we find
 \begin{equation}
 \begin{aligned}
     &G(t)+q_1\partial_t^2G(t)-q_2\partial_t^4G(t)=0\\
     &\partial_t G(0)=G_0=1;\quad\partial_t^nG(0)=(-i)^n\langle H^n\rangle\\
     &q_1=\frac{t^2+2 p^2 v^2}{p^4 v^4};\quad q_2=\frac{1}{p^4 v^4}
     \end{aligned},
 \end{equation}
which can be solved relatively easily to obtain
\begin{equation}
\begin{aligned}
    G(t)=&G_0\frac{\left(1-q_1 \omega_+^2\right) \cos (\omega_- t)+\cos (\omega_+ t)}{\lambda  \omega_+^2}\\
    &-i\langle H\rangle\frac{ \left(\omega_+-q_1 \omega_+^3\right) \sin (\omega_- t)+\omega_- \sin (\omega_+ t)}{\lambda  \omega_- \omega_+^3}\\
    &-\langle H^2\rangle q_2\frac{  \cos (\omega_+ t)-\cos (\omega_- t)}{\lambda }\\
    &+i\langle H^3\rangle q_2\frac{ \omega_- \sin (\omega_+ t)-\omega_+ \sin (\omega_- t)}{\lambda  \omega_- \omega_+}
    \end{aligned},
\end{equation}
 where we made use of shorthand notations
 \begin{equation}
     \omega_\pm=\frac{\sqrt{q_1\mp\lambda}}{\sqrt{2} \sqrt{q_2}};\quad \lambda=\sqrt{q_1^2-4 q_2}.
 \end{equation}
 It has to be stressed that by the obvious replacements
 \begin{equation}
     G_0\to\langle\psi|\phi\rangle;\quad \langle H^n\rangle\to \langle \psi|H^n|\phi\rangle
 \end{equation}
 one obtains transition amplitudes between states $\phi$ and $\psi$.\\
 In summary, this section of the paper lucidly demonstrates how the differential equations for the Loschmidt echo have value not just as an approximate numerical method but can also be used to obtain relatively compact closed-form analytical results in various cases.
 \subsubsection{Fermion Hamiltonian}
 As a next step, we would like to demonstrate the usefulness of this approach for many-body problems. Here, we first consider the case of an interacting fermionic system. We first consider one of the most popular toy problems - the spin-less 1D Fermi-Hubbard chain model. This problem includes a nearest neighbor density interaction and is given below \cite{PhysRevResearch.5.043175}
\begin{equation}
    H=\gamma\sum_{i=1}^L (c_i^\dag c_{i+1}+c_i^\dag c_{i-1})+ U\sum_{i=1}^Lc_i^\dag c_ic_{i+1}^\dag c_{i+1},
    \label{eq:fermi_hubb}
\end{equation}
 where $\gamma$ is the hopping strength of nearest neighbor hoppings, $U$ is the strength of interaction between electrons on neighboring lattice sites, and $L$ is the number of chain sites. Furthermore, we will assume periodic boundary conditions.
 
 Of course, the most important quantity to compute for setting up our differential equation approach is finding expressions for $\mathrm{tr}(H^n)$. To facilitate the computation, we make note of the following identity
 \begin{equation}
     \mathrm{tr}\left(\prod_{j=1}^L(c_j^\dag )^{M_j}c_j^{N_j}\right)=\prod_{j=1}^L \delta_{N_j,M_j}(\delta_{N_j,0}+\delta_{N_j,1})2^{\delta_{N_j,0}},
 \end{equation}
which allows us to compute the traces. Appendix \ref{app:traces_fermi} shows how this identity was derived. It is then easy to see (since we deal with sums) that, generally, the trace will have the form
\begin{equation}
    \mathrm{tr}(H^n)=2^{L-2n}\sum_{j=1}^n\sum_{k=0}^n n_{jk}\gamma^kU^{n-k}L^j,
\end{equation}
where $n_{jk}\in \mathbb{Z}$ are integers. Explicit equations for powers of $n=1\dots 8$ are given in appendix \ref{app:traces_fermi}.\\
Next, we want to demonstrate what happens when we apply our method to this specific kind of Hamiltonian. For this demonstration, we will compute the Loschmidt echo for the ground state $\psi_0$ of a Bogoliubov Hamiltonian \cite{bogoljubov1958new}
\begin{equation}
    H_B=\gamma_0\sum_{i=1}^L (c_i^\dag c_{i+1}+c_i^\dag c_{i-1})+b\sum_i(c_i^\dag c_{i+1}^\dag+c_{i+1}+c_i),
    \label{eq:Hbogo}
\end{equation}
where $b$ is a Bogoliubov coupling. We introduced this coupling to ensure that the ground state has contributions from different particle number sectors of the fermionic Hilbert space. It ensures that our evolution has to perform well on the full Hilbert space - not just on a specific particle number sector - if we are to obtain a reliable result for the Loschmidt echo. It, therefore, ensures more stringent test criteria.\\

\begin{figure*}[htbp!]
    \centering
    \includegraphics[width=1\linewidth]{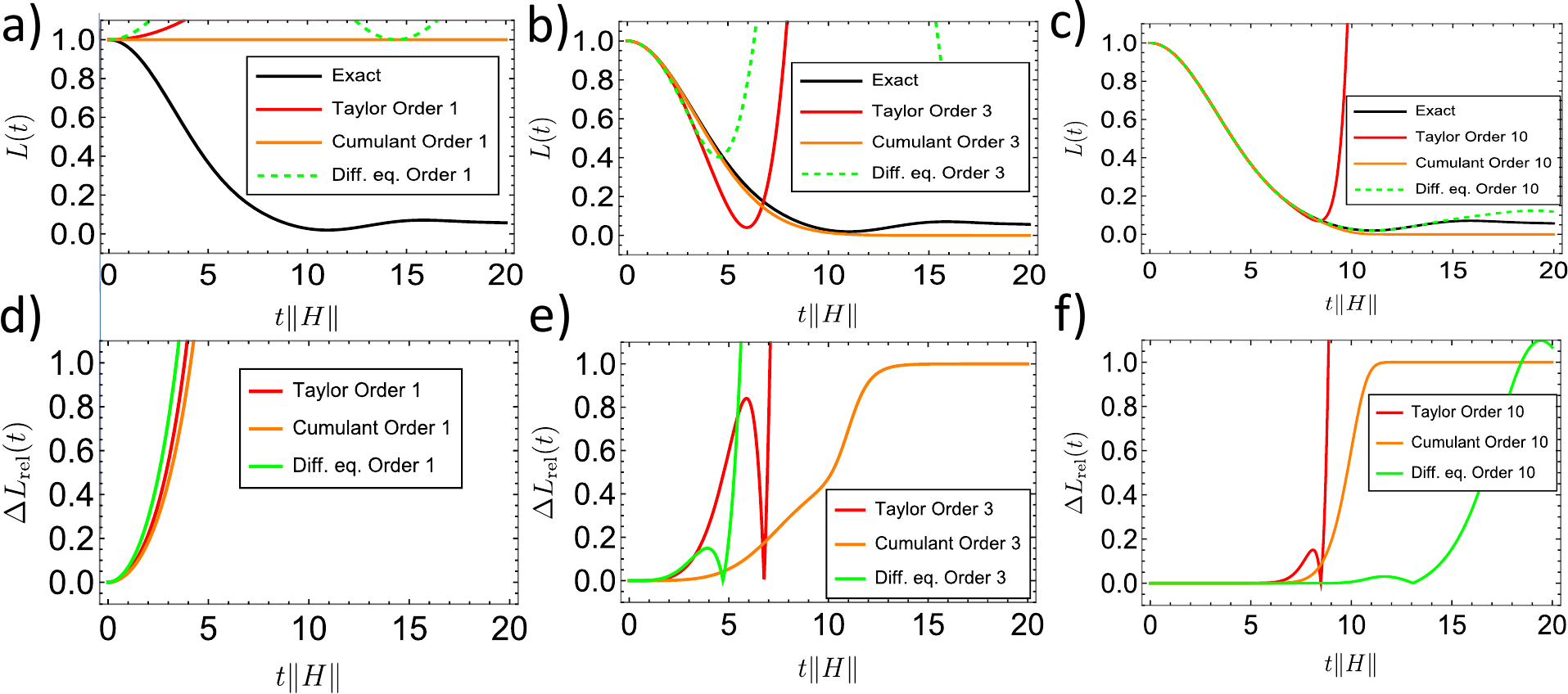}
    \caption{In this figure, we present plots of the Loschmidt echo for a quench in a fermionic system. The starting point is the ground state of the Bogoliubov Hamiltonian $H_B$ in Eq. \eqref{eq:Hbogo} with $\gamma_0=b=1$. The lattice is taken to be a 10-site lattice to have a compromise between numerical speed, interesting features, and the size of Hilbert space. At time $t=0$, the Hamiltonian is quenched to the Fermi-Hubbard model in Eq. \eqref{eq:fermi_hubb} with $\gamma=1$ and $U=3$. The upper row (a-c) shows the Loschmidt as a function of dimensionless time $t\lVert H\rVert$, and the lower row shows the relative mismatch between the exact result and various approximations. In all plots, the exact result \eqref{eq:Loschmidt_amplitude} is plotted in black, the Taylor series result (Eq. \eqref{eq:Taylor}) in red, the result from a cumulant series (Eq. \eqref{eq:Cumulant}) in orange and the result from the differential equation approach (Eq. \eqref{eq:Diffeq}) in green. We note that exact numeric results for the exact diagonalization portion were computed in the Python package QuSpin \cite{weibe1,weibe2}.}
    \label{fig:f5}
\end{figure*}

In the following, we will see that our method cannot be applied blindly - even our previous generic results might leave that impression - but one has to use some insight. To demonstrate this point best, we will first focus on a case where our approach could be better - in that one has to go to relatively large orders (order six or more) to benefit from it. The results from a simple quench can be seen in Fig. \ref{fig:f5}
From this plot, we observe that at low orders, both the cumulant expansion and Taylor expansion are much more reliable than the result from our new method. Regrettably, one has to go to a relatively high order of 6 (not part of the plot) and above to get a result where our method offers benefits, like in the case of order 10. This observation could be seen as a partial failure of the method if we did not obtain a better understanding.\\
Indeed, our approach tries to approximate the Hamiltonian by a linear combination of its higher powers. This approach, of course, only offers significant benefits if pieces of the Hamiltonian have a structure that repeats as we take them to higher powers. Fermionic operators with $ c_i^2=0 $ are a poor choice for such an approach since they generically do not have a repeating structure as we take higher powers (they square to zero). This property makes them the worst-case scenario for studying blindly. 

\begin{figure*}[htbp!]
    \centering
    \includegraphics[width=1\linewidth]{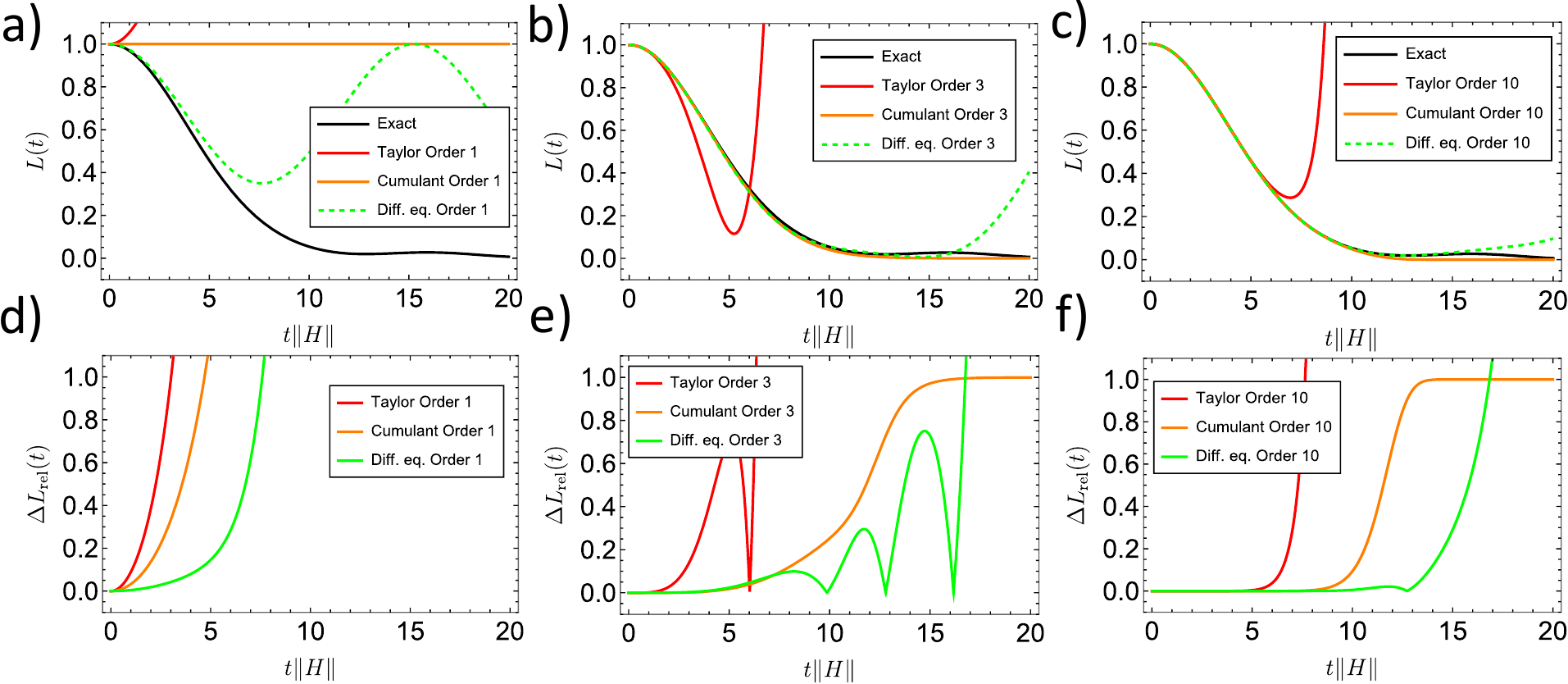}
    \caption{In this figure, we present plots of the Loschmidt echo for a quench in a fermionic system. The starting point is the ground state of the Bogoliubov Hamiltonian $H_B$ in Eq. \eqref{eq:Hbogo} with $\gamma_0=b=1$. The lattice is taken to be a 10-site lattice to have a compromise between numerical speed, interesting features, and the size of Hilbert space. At time $t=0$, the Hamiltonian is quenched to the Fermi-Hubbard model in Eq. \eqref{eq:fermi_hubb} with $\gamma=0.1$ and $U=3$. The upper row (a-c) shows the Loschmidt as a function of dimensionless time $t\lVert H\rVert$, and the lower row shows the relative mismatch between the exact result and various approximations. In all plots, the exact result \eqref{eq:Loschmidt_amplitude} is plotted in black, the Taylor series result (Eq. \eqref{eq:Taylor}) in red, the result from a cumulant series (Eq. \eqref{eq:Cumulant}) in orange and the result from the differential equation approach (Eq. \eqref{eq:Diffeq}) in green.  We note that exact numeric results for the exact diagonalization portion were computed in the Python package QuSpin \cite{weibe1,weibe2}.}
    \label{fig:f6}
\end{figure*}

However, we can analyze in which cases we expect our approach to offer advantages. For our specific Hamiltonian \eqref{eq:fermi_hubb}, the term proportional to $\gamma$ is causing the issue because $(c_i^\dag c_{i+1})^2=0$ - it does not repeat easily. The second term $(c_i^\dag c_ic_{i+1}^\dag c_{i+1})^2=c_i^\dag c_ic_{i+1}^\dag c_{i+1}$ has nicely repeating structure and is therefore unproblematic. This property suggests that we expect the approach to behave well if $t\ll U $. This case is studied in Fig. \ref{fig:f6}.

And we see that even for a first-order approximation, our approach offers a much better approximation than the more standard Cumulant and Taylor expansions.
 \subsubsection{Spin Hamiltonian}
  Next, we want to demonstrate the usefulness of this approach for a many-body spin problem. We consider another very popular toy problem - the Heisenberg model with Dzyaloshinskii–Moriya (DM) interaction -i.e., antisymmetric exchange. This problem includes a nearest neighbor Heisenberg term with strength $J$ and a nearest neighbor antisymmetric exchange term with strength $D$ as given in the Hamiltonian below \cite{PhysRev.120.91}
\begin{equation}
    H=J\sum_{i=1}^L {\bm\sigma}_i\bm \sigma_{i+1}+ D\sum_{i=1}^L(\sigma^x_i\sigma_{i+1}^y-\sigma^y_i\sigma_{i+1}^x),
    \label{eq:spinham}
\end{equation}
where the so-called anisotropy vector of the DM interaction is along the $z$- direction for concreteness. Furthermore, $L$ is the number of chain sites in the expression, and we will assume periodic boundary conditions. To obtain the traces $\mathrm{tr}(H^n)$ necessary to set up differential equations, it is convenient to make use of the following identity
\begin{equation}
    \begin{aligned}
        &\mathrm{tr}\left(\prod_{j=1}^L(\sigma_j^x)^{K_j}(\sigma_j^y)^{M_j}(\sigma_j^z)^{N_j}\right)=\\
        &\prod_{i=1}^L(2\delta_{K_i\mathrm{mod}2+M_i\mathrm{mod}2+N_i\mathrm{mod}2,0}\\
        &+2i\delta_{K_i\mathrm{mod}2+M_i\mathrm{mod}2+N_i\mathrm{mod}2,3})
    \end{aligned},
\end{equation}
which is proved in appendix \ref{app:tracespin}. It is relatively easy to see that traces will typically have the form
\begin{equation}
\mathrm{tr}(H^n) =\sum_{j=1}^{n/2}\sum_{k=1}^n n_{jk}J^kD^{n-k}L^j,
\end{equation}
where $n_{jk}$ are integers. In the same appendix \ref{app:tracespin} we also provide expressions for $\mathrm{tr}(H^n)$ with $n=1,\dots,8$. 

\begin{figure*}[htbp!]
    \centering
    \includegraphics[width=1\linewidth]{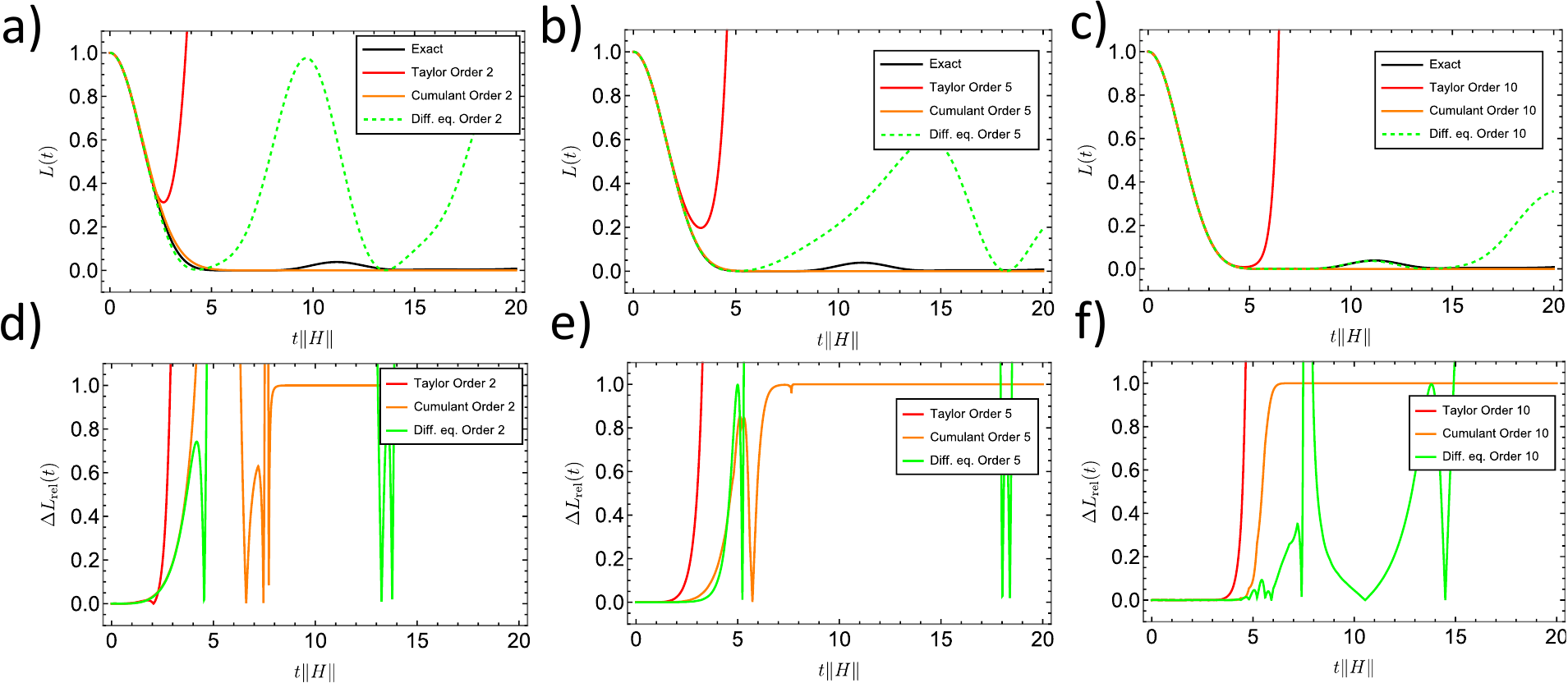}
    \caption{In this figure, we present plots of the Loschmidt echo for a quench in a spin system. The starting point is the ground state of the anti-ferromagnetic Ising chain $H_{\mathrm{Ising}}$ in Eq. \eqref{eq:Hising}. The lattice is taken to be a 10-site lattice to have a compromise between numerical speed, interesting features, and the size of Hilbert space. At time $t=0$, the Hamiltonian is quenched to the Heisenberg chain with DM interaction in Eq. \eqref{eq:spinham} with $J=0.3$ and $D=1$. The upper row (a-c) shows the Loschmidt as a function of dimensionless time $t\lVert H\rVert$, and the lower row shows the relative mismatch between the exact result and various approximations. In all plots, the exact result \eqref{eq:Loschmidt_amplitude} is plotted in black, the Taylor series result (Eq. \eqref{eq:Taylor}) in red, the result from a cumulant series (Eq. \eqref{eq:Cumulant}) in orange and the result from the differential equation approach (Eq. \eqref{eq:Diffeq}) in green.  We note that exact numeric results for the exact diagonalization portion were computed in the Python package QuSpin \cite{weibe1,weibe2}.}
    \label{fig:f7}
\end{figure*}

We need to prepare a state to evolve to test our approach with the spin Hamiltonian. For this, we choose the ground state of an anti-ferromagnetic Ising model that is given as 
\begin{equation}
    H_{\mathrm{Ising}}=\sum_i\left(\sigma_i^z\sigma_{i+1}^z-0.1(-1)^i\sigma_i^z\right),
    \label{eq:Hising}
\end{equation}
A small staggered magnetic field was added as a second term to ensure that ground state degeneracy is broken so that one can choose one consistent ground state. Fig. \ref{fig:f7} presents a representative plot for how different approximations do.
Our approximation scheme offers the advantage of capturing a revival that appears at dimensionless time $t\lVert H\rVert\approx 11$. This feature is completely missed by the cumulant and Taylor series expansions. Regrettably, one has to go to relatively high orders to capture it. 
\section{Estimate for a breakdown time}
Like with any new theoretical tool it is important to have an estimate for when the approximation breaks down. Particularly the approximation error can characterized by
\begin{equation}
    \Delta u=\norm{U_{\text{approx}}-U},
\end{equation}
where $U$ is the exact time evolution operator and $U_{\text{approx}}$ the approximate time evolution operator and $\norm{.}$ an operator norm. This quantity for the purposes of a Loschmidt amplitude is especially useful if we employ a spectral norm. This is because $\langle\psi|A|\psi\rangle\leq \norm{A}$ for a spectral norm and therefore we find a bound on the deviation in Loschmidt amplitude. We will therefore focus on spectral norms.\\
In our case the approximation is controlled by an effective Hamiltonian $H_{\text{approx}}$. In particular for our case, therefore
\begin{equation}
    U=e^{-iHt};\quad U_{\text{approx}}=e^{-iH_{\text{approx}}t}
\end{equation}
Now, if the Hamiltonian $H_{\text{approx}}$ has been estimated to a reasonable order, it should be close to $H$ and we may therefore approximate
\begin{equation}
    U_{\text{approx}}\approx U\left[1-\int dt U^\dag(H-H_{\text{approx}})U\right]
\end{equation}
Using that operator norms can be estimated as $\norm{AB}\leq \norm{A}\norm{B}$ and that the spectral norm of a unitary matrix is 1 we find
\begin{equation}
    \Delta u\lessapprox t \norm{H-H_{\text{approx}}}
\end{equation}
If we want this error to be below a certain cutoff $\epsilon$ we find that the breakdown time $ t_{\text{break}}$ of the approximation can be estimated as
\begin{equation}
    t_{\text{break}}\propto\frac{1}{\norm{H-H_{\text{approx}}}}.
\end{equation}
We notice that the breakdown time is inversely proportional to a spectral norm. Therefore, the breakdown depending on the problem at hand can depend in various ways on the size of the Hilbert space. However, it depends in the same way on the size of the Hilbert space as in the case of a cumulant expansion or a Taylor expansion, which both also have a breakdown time proportional to $1/\norm{H}$. We may observe that $\norm{H-H_{\text{approx}}}$ under many circumstances can be much smaller than $\norm{H}$, which should allow insight into when the different approximations break down,
\section{Conclusion}
\label{sec:5}
To summarize, we have developed a new approximation method that permits us to compute the Loschmidt echo using an approximate differential equation approach. The method is convergent at finite order and, even before it converges, often offers advantages over the more standard Taylor series and cumulant series. In particular, the method can capture revivals in the Loschmidt echo, a critical feature missed by both the Taylor and cumulant expansion. We have also applied the method to some simple band models where obtaining exact results offers some advantages over more straightforward computations - like finding formulas that are easy to generalize for arbitrary starting vectors or transmission amplitudes.

Of course, like with any new approach, there are some drawbacks - some of them might be remedied later. For instance, in some cases, one needs to go to relatively high approximate orders to reap the approach's benefits - is there a way to accelerate convergence? Moreover, additional progress is needed in developing a clearer understanding of the circumstances when the method breaks down in the sense that the cumulant expansion outperforms it - is there a way to avoid this? Moreover, obtaining estimates for the time frame until which results are reliable is an outstanding question that seems difficult to answer. 

Other interesting possible future directions to explore are i) generalizing the approach to obtain differential equations for the partition function ii), Lindblad master equation approaches to the fidelity, which could provide fascinating insights into the stability of qubits iii) there are reasons to believe the approach could be employed in the context of high energy physics to obtain improved scattering matrix elements (interaction strength taking the role of time). Another interesting direction could be to exploit that the series is convergent and make use of convergence acceleration techniques.
\acknowledgments
 M.V. gratefully acknowledges the support provided by the Deanship of Research Oversight and Coordination (DROC) at King Fahd University of Petroleum \& Minerals (KFUPM) for funding this work through exploratory research grant No. ER221002.
\bibliographystyle{unsrt}
\bibliography{literature}

\begin{thebibliography}{10}

\bibitem{Burnett_2019}
Jonathan~J. Burnett, Andreas Bengtsson, Marco Scigliuzzo, David Niepce, Marina
  Kudra, Per Delsing, and Jonas Bylander.
\newblock Decoherence benchmarking of superconducting qubits.
\newblock {\em npj Quantum Information}, 5(1), June 2019.

\bibitem{PhysRevLett.128.127702}
Anne Matthies, Jinhong Park, Erez Berg, and Achim Rosch.
\newblock Stability of floquet majorana box qubits.
\newblock {\em Phys. Rev. Lett.}, 128:127702, Mar 2022.

\bibitem{PhysRevB.109.144502}
Matthew~F. Lapa and Michael Levin.
\newblock Stability of topological superconducting qubits with number
  conservation.
\newblock {\em Phys. Rev. B}, 109:144502, Apr 2024.

\bibitem{roberts2022fidelity}
Wesley Roberts, Michael Vogl, and Gregory~A. Fiete.
\newblock Fidelity of the kitaev honeycomb model under a quench, 2022.

\bibitem{GORIN200633}
Thomas Gorin, Tomaž Prosen, Thomas~H. Seligman, and Marko Žnidarič.
\newblock Dynamics of loschmidt echoes and fidelity decay.
\newblock {\em Physics Reports}, 435(2):33--156, 2006.

\bibitem{Heyl_2018}
Markus Heyl.
\newblock Dynamical quantum phase transitions: a review.
\newblock {\em Reports on Progress in Physics}, 81(5):054001, apr 2018.

\bibitem{vzunkovivc2016dynamical}
Bojan {\v{Z}}unkovi{\v{c}}, Alessandro Silva, and Michele Fabrizio.
\newblock Dynamical phase transitions and loschmidt echo in the infinite-range
  xy model.
\newblock {\em Philosophical Transactions of the Royal Society A: Mathematical,
  Physical and Engineering Sciences}, 374(2069):20150160, 2016.

\bibitem{doi:10.1137/S00361445024180}
Cleve Moler and Charles Van~Loan.
\newblock Nineteen dubious ways to compute the exponential of a matrix,
  twenty-five years later.
\newblock {\em SIAM Review}, 45(1):3--49, 2003.

\bibitem{Heyl_2019survey}
Markus Heyl.
\newblock Dynamical quantum phase transitions: A brief survey.
\newblock {\em EPL (Europhysics Letters)}, 125(2):26001, February 2019.

\bibitem{Tonielli_2019}
F.~Tonielli, R.~Fazio, S.~Diehl, and J.~Marino.
\newblock Orthogonality catastrophe in dissipative quantum many-body systems.
\newblock {\em Physical Review Letters}, 122(4), January 2019.

\bibitem{jacquod2009decoherence}
Ph~Jacquod and Cyril Petitjean.
\newblock Decoherence, entanglement and irreversibility in quantum dynamical
  systems with few degrees of freedom.
\newblock {\em Advances in Physics}, 58(2):67--196, 2009.

\bibitem{Castro_Neto_2009}
A.~H. Castro~Neto, F.~Guinea, N.~M.~R. Peres, K.~S. Novoselov, and A.~K. Geim.
\newblock The electronic properties of graphene.
\newblock {\em Reviews of Modern Physics}, 81(1):109–162, January 2009.

\bibitem{McCann_2013}
Edward McCann and Mikito Koshino.
\newblock The electronic properties of bilayer graphene.
\newblock {\em Reports on Progress in Physics}, 76(5):056503, April 2013.

\bibitem{Min_2008}
Hongki Min and Allan~H. MacDonald.
\newblock Electronic structure of multilayer graphene.
\newblock {\em Progress of Theoretical Physics Supplement}, 176:227–252,
  2008.

\bibitem{PhysRevX.7.021019}
Qihang Liu and Alex Zunger.
\newblock Predicted realization of cubic dirac fermion in quasi-one-dimensional
  transition-metal monochalcogenides.
\newblock {\em Phys. Rev. X}, 7:021019, May 2017.

\bibitem{annurev:/content/journals/10.1146/annurev-conmatphys-031016-025458}
Binghai Yan and Claudia Felser.
\newblock Topological materials: Weyl semimetals.
\newblock {\em Annual Review of Condensed Matter Physics}, 8(Volume 8,
  2017):337--354, 2017.

\bibitem{Bouhlal_2022}
Ahmed Bouhlal, Adel Abbout, Ahmed Jellal, Hocine Bahlouli, and Michael Vogl.
\newblock Tunneling phase diagrams in anisotropic multi‐weyl semimetals.
\newblock {\em Annalen der Physik}, 534(11), September 2022.

\bibitem{PhysRevLett.108.140405}
S.~M. Young, S.~Zaheer, J.~C.~Y. Teo, C.~L. Kane, E.~J. Mele, and A.~M. Rappe.
\newblock Dirac semimetal in three dimensions.
\newblock {\em Phys. Rev. Lett.}, 108:140405, Apr 2012.

\bibitem{PhysRevResearch.6.013048}
Arpit Raj, Swati Chaudhary, and Gregory~A. Fiete.
\newblock Photogalvanic response in multi-weyl semimetals.
\newblock {\em Phys. Rev. Res.}, 6:013048, Jan 2024.

\bibitem{Fang_2012}
Chen Fang, Matthew~J. Gilbert, Xi~Dai, and B.~Andrei Bernevig.
\newblock Multi-weyl topological semimetals stabilized by point group symmetry.
\newblock {\em Physical Review Letters}, 108(26), June 2012.

\bibitem{BOUHLAL2021168563}
Ahmed Bouhlal, Ahmed Jellal, Hocine Bahlouli, and Michael Vogl.
\newblock Tunneling in an anisotropic cubic dirac semi-metal.
\newblock {\em Annals of Physics}, 432:168563, 2021.

\bibitem{PhysRevResearch.5.043175}
Qingyu Li, Chiranjib Mukhopadhyay, and Abolfazl Bayat.
\newblock Fermionic simulators for enhanced scalability of variational quantum
  simulation.
\newblock {\em Phys. Rev. Res.}, 5:043175, Nov 2023.

\bibitem{bogoljubov1958new}
Nikolay~N Bogoljubov, Vladimir~Veniaminovic Tolmachov, and DV~{\v{S}}irkov.
\newblock A new method in the theory of superconductivity.
\newblock {\em Fortschritte der physik}, 6(11-12):605--682, 1958.

\bibitem{weibe1}
Phillip Weinberg and Marin Bukov.
\newblock {QuSpin: a Python package for dynamics and exact diagonalisation of
  quantum many body systems part I: spin chains}.
\newblock {\em SciPost Phys.}, 2:003, 2017.

\bibitem{weibe2}
Phillip Weinberg and Marin Bukov.
\newblock {QuSpin: a Python package for dynamics and exact diagonalisation of
  quantum many body systems. Part II: bosons, fermions and higher spins}.
\newblock {\em SciPost Phys.}, 7:020, 2019.

\bibitem{PhysRev.120.91}
T\^oru Moriya.
\newblock Anisotropic superexchange interaction and weak ferromagnetism.
\newblock {\em Phys. Rev.}, 120:91--98, Oct 1960.

\bibitem{PhysRevB.104.184425}
Utkarsh Bajpai, Abhin Suresh, and Branislav~K. Nikoli\ifmmode~\acute{c}\else
  \'{c}\fi{}.
\newblock Quantum many-body states and green's functions of nonequilibrium
  electron-magnon systems: Localized spin operators versus their mapping to
  holstein-primakoff bosons.
\newblock {\em Phys. Rev. B}, 104:184425, Nov 2021.

\end{thebibliography}

\appendix
\section{Trace identities for fermions}
\label{app:traces_fermi}
In this section, we will compute some of the trace identities that are quoted in the main text. We first start with the following trace
\begin{equation}
T=\mathrm{tr}\left(\prod_{j=1}^L(c_j^\dag )^{M_j}c_j^{N_j}\right)
\end{equation}
where $M_j,N_k\in \mathbb{N}_0$. We may now use the Jordan-Wigner transform for a lattice with $L$ sites in a tensor product form\cite{PhysRevB.104.184425}
\begin{equation}
    c_j^\dag=\bigotimes_{k=1}^{j-1}\sigma_z^{(k)}\otimes\sigma_+^{(j)}\bigotimes_{l=j+1}^{L}\mathbb{1}_2^{(l)},
\end{equation}
where we put superscripts in parentheses that denote lattice sites.
We find
\begin{equation}
T=\mathrm{tr}\left(\prod_{j=1}^L\left(\bigotimes_{k=1}^{j-1}\mathbb{1}_2^{(k)}\otimes (\sigma_+^{(j)})^{M_j}(\sigma_-^{(j)})^{N_j}\bigotimes_{l=j+1}^L\mathbb{1}_2^{(k)}\right)\right),
\end{equation}
 We may  simplify the product further to obtain
\begin{equation}
T=\mathrm{tr}\left(\bigotimes_{j=1}^L(\sigma_+^{(j)})^{M_j}(\sigma_-^{(j)})^{N_j}\right).
\end{equation}
The trace of a tensor product can now be factorized as
\begin{equation}
T=\prod_{j=1}^L\mathrm{tr}\left((\sigma_+^{(j)})^{M_j}(\sigma_-^{(j)})^{N_j}\right).
\end{equation}
Now it is important to note that each of the traces is only non-zero if $M_j=N_j$ such that we find
\begin{equation}
    T=\prod_{j=1}^L \delta_{N_j,M_j}\mathrm{tr}\left((\sigma_+^{(j)})^{M_j}(\sigma_-^{(j)})^{M_j}\right)
\end{equation}
Next, we note that
\begin{equation}
    \mathrm{tr}\left(\sigma_+^{M}\sigma_-^{M}\right)=\begin{cases}
    2;\quad M=0\\
    1;\quad M=1\\
    0;\quad\text{otherwise}\\
    \end{cases},
\end{equation}
which allows us to find 
\begin{equation}
    T=\prod_{j=1}^L \delta_{N_j,M_j}(\delta_{N_j,0}+\delta_{N_j,1})2^{\delta_{N_j,0}}.
\end{equation}

This type of expression may now be used to compute traces $\mathrm{tr}(H^n)$. Here, it is useful to start with smaller system sizes and take note that
\begin{equation}
    \mathrm{tr}(H^n)/(2^{L-2n})=\sum_{j,k=1}^n n_{jk}\gamma^kU^{n-k}L^j,
\end{equation}
where $n_{jk}$ are integers. This equation also implies that if $U,\gamma$ are chosen as integers, the coefficients in front of $L^j$ are integers. That is, for $U,\gamma\in \mathbb{N}$ we have
\begin{equation}
    \mathrm{tr}(H^n)/(2^{L-2n})=\sum_{j,k=1}^n \mathcal{N}_j(\gamma\in\mathbb{Z},U\in\mathbb{Z})L^j ,
\end{equation}
where $\mathcal{N}_j(\gamma\in\mathbb{Z},U\in\mathbb{Z})\in\mathbb{Z}$.

For each pair of $(\gamma,U)$ one then needs only to match $n$  coefficients - that is consider $n$ different lattice size results - to obtain a general expression valid for arbitrary lattice size $L$ - a least square-fit can do this. One should note that at order $n$, one has to consider lattice sizes of $L=n+1$ and larger to avoid debilitating finite size effects. 

To then obtain the $\gamma$ and $U$ dependence one then has to find $n$ non-trivial combinations of $(\gamma,U)$ and fit the $n_{jk}$ such that
\begin{equation}
    \mathcal{N}_j=\sum_kn_{jk}\gamma^kU^{n-k}.
\end{equation}
Using this approach, we found the following identities below
\begin{widetext}
\begin{equation}
    \mathrm{tr}(H^0)=2^L;\quad \mathrm{tr}(H^1)=2^{L-2} L U
\end{equation}
\begin{equation}
    \mathrm{tr}(H^2)=2^{L-4} \left[L \left(8 \gamma^2+5 U^2\right)+L^2 U^2\right]
\end{equation}
\begin{equation}
\begin{aligned}
    \mathrm{tr}(H^3)=&2^{L-6} \left[L \left(24 U^3-24 \gamma^2 U\right)+L^2 \left(24 \gamma^2 U+15 U^3\right)+L^3 U^3\right]
    \end{aligned}
\end{equation}
\begin{equation}
\begin{aligned}
    \mathrm{tr}(H^4)=&2^{L-8} \left[L \left(62 U^4-192 \gamma^4-448 \gamma^2 U^2\right)+L^2 \left(192 \gamma^4+144 \gamma^2 U^2+171 U^4\right)+L^3 \left(48 \gamma^2 U^2+30 U^4\right)+L^4 U^4\right]
    \end{aligned}
\end{equation}
\begin{equation}
\begin{aligned}
    \mathrm{tr}(H^5)=&2^{L-10} \left[L \left(3840 \gamma^4 U-3360 \gamma^2 U^3-960 U^5\right)+L^2 \left(-2880 \gamma^4 U-1520 \gamma^2 U^3+1510 U^5\right)\right.\\
    &\left.+L^3 \left(960 \gamma^4 U+960 \gamma^2 U^3+615 U^5\right)+L^4 \left(80 \gamma^2 U^3+50 U^5\right)+L^5 U^5\right]
    \end{aligned}
\end{equation}
\begin{equation}
\begin{aligned}
    \mathrm{tr}(H^6)=&2^{L-12} \left[L \left(20480 \gamma^6+90240 \gamma^4 U^2+24192 \gamma^2 U^4-22960 U^6\right)\right.\\
    &\left.+L^2 \left(4650 U^6-23040 \gamma^6-39360 \gamma^4 U^2-57840 \gamma^2 U^4\right)+L^3 \left(7680 \gamma^6+6600 \gamma^2 U^4+10005 U^6\right)\right.\\
    &\left.+L^4 \left(2880 \gamma^4 U^2+3120 \gamma^2 U^4+1605 U^6\right)+L^5 \left(120 \gamma^2 U^4+75 U^6\right)+L^6 U^6\right]
    \end{aligned}
\end{equation}
\begin{equation}
\begin{aligned}
    \mathrm{tr}(H^7)=&2^{L-14} \left[L \left(1436288 \gamma^2 U^5-1182720 \gamma^6 U+297472 \gamma^4 U^3-233856 U^7\right)\right.\\
    &\left.+L^2 \left(949760 \gamma^6 U+685440 \gamma^4 U^3-773136 \gamma^2 U^5-209440 U^7\right)\right.\\
    &\left.+L^3 \left(115710 U^7-322560 \gamma^6 U-396480 \gamma^4 U^3-195720 \gamma^2 U^5\right)\right.\\
    &\left.+L^4 \left(53760 \gamma^6 U+53760 \gamma^4 U^3+62440 \gamma^2 U^5+40495 U^7\right)\right.\\
    &\left.+L^5 \left(6720 \gamma^4 U^3+7560 \gamma^2 U^5+3465 U^7\right)+L^6 \left(168 \gamma^2 U^5+105 U^7\right)+L^7 U^7\right]
    \end{aligned}
\end{equation}
\begin{equation}
\begin{aligned}
    \mathrm{tr}(H^8)=&2^{L-16} \left[L \left(1736432 U^8-4874240 \gamma^8-28442624 \gamma^6 U^2-41072640 \gamma^4 U^4+26372096 \gamma^2 U^6\right)\right.\\
    &\left.+L^2 \left(5877760 \gamma^8+14479360 \gamma^6 U^2+36300544 \gamma^4 U^4+4564224 \gamma^2 U^6-6240948 U^8\right)\right.\\
    &\left.+L^3 \left(99260 U^8-2580480 \gamma^8-931840 \gamma^6 U^2-6518400 \gamma^4 U^4-8495424 \gamma^2 U^6\right)\right.\\
    &\left.+L^4 \left(430080 \gamma^8-860160 \gamma^6 U^2-846720 \gamma^4 U^4+285600 \gamma^2 U^6+807345 U^8\right)\right.\\
    &\left.+L^5 \left(215040 \gamma^6 U^2+282240 \gamma^4 U^4+260960 \gamma^2 U^6+124040 U^8\right)\right.\\
    &\left.+L^6 \left(13440 \gamma^4 U^4+15456 \gamma^2 U^6+6594 U^8\right)+L^7 \left(224 \gamma^2 U^6+140 U^8\right)+L^8 U^8\right]
    \end{aligned}
\end{equation}
\end{widetext}
It is important to stress that the results for $\mathrm{tr}(H^n)$ are exactly valid for chains of an arbitrary number of sites $L>n$. They are not valid for $L<n+1$ because the periodic boundary conditions spoil the result for a smaller number of sites $L$.
\section{Trace identities for spins}
\label{app:tracespin}
In this section, we will compute some trace identities for spin Hamiltonians that are just quoted in the main text. First, we compute the following useful general trace
\begin{equation}
    T=\mathrm{tr}\left(\prod_{j=1}^L(\sigma_j^x)^{K_j}(\sigma_j^y)^{M_j}(\sigma_j^z)^{N_j}\right),
\end{equation}
where $K_j,M_j,N_j\in \mathbb{Z}$. Making use of the fact that the spin operators can be rewritten as tensor product \cite{PhysRevB.104.184425}
\begin{equation}
    \sigma_j^{x,y,z}=\bigotimes_{k=1}^{j-1}\mathbb{1}_2^{(k)}\otimes \sigma_{x,y,z}^{(j)}\bigotimes_{l=j+1}^{L}\mathbb{1}_2^{(l)}.
\end{equation}
For clarity, we introduced a site label in parentheses. We then obtain
\begin{equation}
    T=\mathrm{tr}\left(\bigotimes_{i=1}^L \left[(\sigma_i^x)^{K_i}(\sigma_i^y)^{M_i}(\sigma_i^z)^{N_i}\right]^{(i)}\right)
\end{equation}
Now, using the trace of a tensor product factories, we find that
\begin{equation}
    T=\prod_{i=1}^L\mathrm{tr}\left((\sigma_i^x)^{K_i}(\sigma_i^y)^{M_i}(\sigma_i^z)^{N_i}\right)
\end{equation}
The result can be simplified further by making use of the fact that Pauli matrices square to the identity matrix to obtain
\begin{equation}
    T=\prod_{i=1}^L\mathrm{tr}\left((\sigma_i^x)^{K_i\mathrm{mod}2}(\sigma_i^y)^{M_i\mathrm{mod}2}(\sigma_i^z)^{N_i\mathrm{mod}2}\right)
\end{equation}
Now, there is only two combinations of exponents that lead to a non-zero trace $K_i\mathrm{mod}2=M_i\mathrm{mod}2=N_i\mathrm{mod}2=1$ and $K_i\mathrm{mod}2=M_i\mathrm{mod}2=N_i\mathrm{mod}2=0$. Taking these cases into consideration, one obtains
\begin{equation}
    \begin{aligned}
        T=&\prod_{i=1}^L(2\delta_{K_i\mathrm{mod}2+M_i\mathrm{mod}2+N_i\mathrm{mod}2,0}\\
        &+2i\delta_{K_i\mathrm{mod}2+M_i\mathrm{mod}2+N_i\mathrm{mod}2,3})
    \end{aligned}.
\end{equation}

It is now straightforward to compute traces $\mathrm{tr}(H^n)$ of the spin Hamiltonian \eqref{eq:spinham}. Especially if we note that the general form the result we will obtain in this particular case has the general form
\begin{equation}
\mathrm{tr}(H^n) =\sum_{j=1}^{n/2}\sum_{k=1}^n n_{jk}J^kD^{n-k}L^j,
\end{equation}
where $n_{jk}\in \mathbb{Z}$

Results for the first few orders are given below
\begin{widetext}
    \begin{equation}
        \mathrm{tr}(H)=0;\quad \mathrm{tr}(H^2)=2^L L \left(2 D^2+3 J^2\right);\quad \mathrm{tr}(H^3)=-2^L L \left(6 D^2 J+6 J^3\right)
    \end{equation}
    \begin{equation}
        \mathrm{tr}(H^4)=2^L \left[L^2 \left(12 D^4+36 D^2 J^2+27 J^4\right)-L \left(12 D^4+40 D^2 J^2+30 J^4\right)\right]
    \end{equation}
    \begin{equation}
        \mathrm{tr}(H^5)=15\ 2^{L+2} \left[L \left(4 D^4 J+10 D^2 J^3+6 J^5\right)-L^2 \left(2 D^4 J+5 D^2 J^3+3 J^5\right)\right]
    \end{equation}
      \begin{equation}
      \begin{aligned}
          \mathrm{tr}(H^6)=&2^L \left[L^3 \left(120 D^6+540 D^4 J^2+810 D^2 J^4+405 J^6\right)\right.\\&\left.+L^2 \left(-360 D^6-1380 D^4 J^2-1980 D^2 J^4-990 J^6\right)\right.\\&\left.+L \left(320 D^6+840 D^4 J^2+1008 D^2 J^4+504 J^6\right)\right]
      \end{aligned}
    \end{equation}
     \begin{equation}
      \begin{aligned}
          \mathrm{tr}(H^7)=&2^{L+1} \left[L^2 \left(6300 D^6 J+25620 D^4 J^3+33810 D^2 J^5+14490 J^7\right)\right.\\&\left.-L \left(9240 D^6 J+38584 D^4 J^3+51352 D^2 J^5+22008 J^7\right)\right.\\&\left.-L^3 \left(1260 D^6 J+5040 D^4 J^3+6615 D^2 J^5+2835 J^7\right)\right]
      \end{aligned}
    \end{equation}
         \begin{equation}
      \begin{aligned}
          \mathrm{tr}(H^8)=&2^L \left[L^4 \left(1680 D^8+10080 D^6 J^2+22680 D^4 J^4+22680 D^2 J^6+8505 J^8\right)\right.\\&\left.-L^3 \left(10080 D^8+43680 D^6 J^2+78120 D^4 J^4+70560 D^2 J^6+26460 J^8\right)\right.\\&\left.+L^2 \left(22960 D^8+26880 D^6 J^2-74032 D^4 J^4-125664 D^2 J^6-47124 J^8\right)\right.\\&\left.+L \left(204048 J^8-19040 D^8+88704 D^6 J^2+447552 D^4 J^4+544128 D^2 J^6\right)\right],
      \end{aligned}
    \end{equation}
\end{widetext}
where much like the fermionic case our expressions for $\mathrm{tr}(H^n)$ are valid for arbitrary number of sites $L>n$. The boundary conditions spoil the case for a smaller number of sites.
\end{document}